\documentclass[aps,pra,preprint,groupedaddress,amsmath,amssymb,showpacs,showkeys]
{revtex4-1}
\usepackage[dviwin]{graphicx}
\usepackage[dvips]{color}
\usepackage{graphicx}
\usepackage{dcolumn}
\usepackage{bm}
\begin{document}
%
%
%
\title{Visualization of Rabi oscillations  in a magnetic resonance}
%

\author{E. A. Ivanchenko}
\email{eaivanchenko1@gmail.com,yevgeny@kipt.kharkov.ua}

\affiliation{Institute for Theoretical Physics, National Science
Center \textquotedblleft{}Institute of Physics and
Technology\textquotedblright{},
 \\
  1, Akademicheskaya
str., 61108 Kharkov, Ukraine}
\date{\today}
\begin{abstract}
On the basis of solutions of the equations for a density matrix in Hermitian base
the scheme of visualization of dynamics for the polarization vector of  qudit in a time-dependent magnetic field  is presented by means of mapping of the solution for the polarization vector  on the three-dimensional spherical curve (vector hodograph).
      The received results obviously display the interference  of precessional and   nutational effects on  the polarization vector  in a magnetic resonance. The study can find  practical applications in magnetic resonance and  the 3D visualization of computational data.
\\
 \\
 \end{abstract}



\maketitle
%
\section{Introduction}\label{sec:1}
Imaging and visualization are among the most dynamic and
innovative  research areas of  the past  decades.
 This activity arises from the requirements of
important practical applications such as the computational data
visualization, the medical images processing   for assisting
medical diagnosis and intervention, and the 3D geometry
reconstruction and processing for computer simulations.
Due to the development of more powerful hardware
resources, mathematical and physical methods, investigators have
been incorporating advanced computational techniques to derive
 methodologies that can better enable the solution of
the problems encountered.\\
\indent In this paper we are going to introduce the 3D visualization scheme   of the qudit polarization vector evolution  in a magnetic field.
As it is known, the magnetic resonance realization depends on the kind of the magnetic field modulations. Let's consider the spin dynamics in an alternating
field \cite{PhysicaBIvanchenko,StabilizationIvanchenko}
\begin{equation}\label{eq:1}
\overrightarrow {h (t)}=\left( h_1 \mathrm{cn}(\omega t|k),\;h_2 \mathrm{sn}
(\omega t|k),\;H \mathrm{dn}(\omega t|k)\right),
\end{equation}
where $\mathrm{cn},\mathrm{sn},\mathrm{dn}$ are the Jacobi
elliptic functions \cite{AbramovitzStegun}, $\omega$ is the field frequency. Such field modulation
under the changing of the elliptic modulus $k$ from 0 to 1
describes the whole class of field forms from
trigonometric \cite{IIRabi} ($\mathrm{cn}(\omega
t|0)=\mathrm{cos}\omega t,\;\mathrm{sn}(\omega
t|0)=\mathrm{sin}\omega t,\;\mathrm{dn}(\omega t|0)=1$)  to the
exponentially impulse ones ($\mathrm{cn}(\omega t|1)=
\frac{1}{\mathrm{ch}\omega t},\;\mathrm{sn}(\omega
t|1)=\mathrm{th}\omega t,\;\mathrm{dn}(\omega
t|1)=\frac{1}{\mathrm{ch}\omega t}$) \cite{BambiniBerman}. The
elliptic functions $\mathrm{cn}(\omega t|k)$ and$\;
\mathrm{sn}(\omega t|k)$ have the real period $\frac{4K}{\omega
}$, while the function $\mathrm{dn}(\omega t|k)$ has a period of
half the duration. Here $K$ is the full elliptic integral of the
first kind \cite{AbramovitzStegun}.
We call such field consistent.\\
\indent At $k=0$ and $h_1=h_2=h $ it is a circularly polarized magnetic field
 \cite {IIRabi}, where the scalar  of the vector of the magnetic field does not depend on the time (a rotating magnetic field), and at $k=0$ and
$h_1=h, h_2=0$ it is a linearly polarized field \cite{BlochSiegert}.\\
\indent The solution of the Schrodinger equation for a wave function does not yield
directly to  the physical observed values. In the paper
\cite{FeynmanVernonHellwarth} real  functions have been constructed  from  the Schrodinger equation solutions, which have direct physical sense and
whose temporary evolution supposes visualization.
  The objective of this work is  to present the   visualization scheme of   qudit polarization vector dynamics   in a magnetic field on the base of
both analytical and numerical
solutions for the density matrix in the Hermitian base.
The solution represents a real generalized Bloch vector. The 3D
sphere is used for the  numerical simulation results   visualization.
 The solution of the equations of motion for the polarization vector maps on tri-dimensional oriented spherical
curve, which in the case of qudits shows a precession similar to that of a symmetric top in the gravitation field \cite{Goldstein}.\\
\indent This paper  is organized as follows.
Section 2 describes in the presentation of the real Bloch vector a set of equations for qubit dynamics taking into account its  environment.
In Section 3 we present the  3D visualization scheme of the qudit polarization vector.
 On the base of the analytic solution in a rotating magnetic field the scheme of  the qubit polarization vector visualization  and the analytical formulas  characterizing  visualization in a resonance case,
 are presented in Section 4.
  Section 5 contains the analytical solution for qutrit  in a rotating magnetic field in case of  resonance, taking into account the anisotropy, as an example   of the qudit polarization vector visualization.
   In Section 6 the results are presented pictorially at concrete parameters.  The deductions are presented in the short conclusion.  In the Appendix  the subsidiary analytical results are added.
\section{Hamiltonian and master equation in Lindblad form}\label{sec:2}
The Hamiltonian of a magnetic qubit (spin 1/2 particle)
which is in the external variable magnetic field
$ \overrightarrow {h (t)} = (h_1 (t), h_2 (t), h_3 (t)) $  is equal to
\begin{equation}\label{eq:2}
\hat{H}=h_i(t)s_i,
\end{equation}
where $h_i (t) $ are the Cartesian components of the external magnetic field in frequency units (we suppose $ \hbar=1$);
 $s_i =\frac {1} {2} \sigma_i $,
 $\sigma_i $ are the  Pauli matrices.
\\
\indent The  Liouville - Neumann equation for the density matrix $ \rho $,
describing the qubit dynamics and taking into account the  environment in the form of Lindblad, takes the form
\begin{equation} \label {eq:3}
\partial_t\rho =-i [\hat {H}, \rho] +
\frac {1} {2} \sum^3 _ {\alpha, \beta=1}
a _ {\alpha\beta} (2\sigma_\alpha\rho\sigma_\beta
-\sigma_\beta\sigma_\alpha\rho-\rho\sigma_\beta\sigma_\alpha), ~
\rho (t=0) = \rho_0,
\end{equation}
where $a _ {11} =a _ {22} = \gamma_1/4, a _ {33} = \gamma_2/2-\gamma_1/4,
a _ {12} =a _ {12} ^ * =-i\gamma_1R_3 (t=0)/4, a _ {23} =a _ {32} =a _ {31} =a _ {13} =0$
\cite{FBloch,ViolaFortunatoLloydTsengCory}. The constants
$ \gamma_1$, $ \gamma_2$, and $R _ {eqr} $ are identified
as the longitudinal and transverse lifetimes, and the equilibrium
value of the population difference respectively.
\\
\indent We  present the equation solution 
\begin{equation} \label {eq:4}
  \rho =\frac {1} {2} R _ {\alpha} \sigma_\alpha, ~
  \rho ^ + =\rho, ~ \mathrm {Tr \,} \rho=1, ~R _ {0} =1,
\end{equation}
in which, here and further, we imply the summation on  repetitive Greek
coefficients from zero to three and on Latin from one to three. The coherence vector (the Bloch vector) which is widely used in the theory of the
magnetic resonance,
\begin{equation} \label {eq:5}
R _ {i} = \mathrm {Tr \,} (\rho\sigma_i),
\end{equation}
characterizes the qubit behavior. At unitary evolution
the length of  the Bloch vector $ b $ is conserved
\begin{equation} \label {eq:6}
b = \sqrt {R^2 _ {i}}.
\end{equation}
 In the terms of functions $R _ {i} $
the  Liouville - Neumann equation takes the form of the
closed system of 3 differential equations 
for the Bloch vector components
\begin{eqnarray}\label {eq:7}
\partial_t R _ {1}&=&h_2 R _ {3}-h_3 R _ {2}-\gamma_2 R _ {1}, \\
\partial_t R _ {2}&=&h_3 R _ {1}-h_1 R _ {3}-\gamma_2 R _ {2}, \\
\partial_t R _ {3}&=&h_1 R _ {2}-h_2 R _ {1}-\gamma_1 (
R _ {3}-R _ {eqr})
\end{eqnarray}
 at the initially given conditions.
\section{Visualization scheme of dynamics}\label{sec:3}
One of the major stages of investigations is the visualization of the data obtained.
"As known, the difficulty arising in quantum mechanics is not only to find the solutions but also to understand their meaning" \cite {FeynmanLectures}. In this paper it will be useful to map
the solution for the polarization vector on geometrical
model \cite {FeynmanVernonHellwarth}. To this end, it is convenient to parameterize the unit polarization vector by spherical angles $ (p_1, p_2, p_3) = (\sin\theta
\cos\varphi, \sin\theta \sin\varphi, \cos\theta)$.
Thus, the parameter $\varphi $ (0$ \leq\varphi\leq 2\pi $) becomes a precession angle
of the end of the vector $ \vec {p} $ on a sphere, referred to as
apex, and the angle $ \theta $ (0$ \leq\theta\leq \pi $)
characterizes the nutation. $ \theta=0$ corresponds to the north pole on the  sphere. These angles are expressed through the components of
polarization vector $ \vec {p} $ \, according to the following formulae:
 \begin{equation} \label {eq:10}
\theta (t) = \arccos
p_3, \, \sin\varphi (t) = \frac {p_2} {\sqrt {p_1^2+p_2^2}}, \,
\cos\varphi (t) = \frac {p_1} {\sqrt {p_1^2+p_2^2}}.
\end{equation}
The angular velocities are given by
 \begin{equation} \label {eq:11}
\theta \,' (t) = \frac {h_2 p_1-h_1 p_2} {\sqrt {1-p_3^2}}, \,
\varphi\,'(t) = (\arctan \frac {p_2} {p_1}) '.
\end{equation}
The oriented spherical curve 
is characterized by the curvature $k$, torsion of the
curve $ \kappa $,  velocity modulus of an apex $ \mathrm {v} $, and length
of a path $s$. All these quantities can be easily calculated in terms of the components of polarization vector using the following formulae \cite{Aminov}:
\begin{equation} \label {eq:12}
k =\frac {| [\vec {p} \,', \vec {p}\,''] |} {|\vec {p}\,' | ^3},
\kappa =\frac {(\vec {p}\,', \vec {p}\,'', \vec {p}\,''')} {[\vec {p}\,', \vec {p}\,''] ^ 2},
 \mathrm {v} = \sqrt {p_1'^2+p_2'^2+p_3'^2}, \,
s =\int_0^t \mathrm {v}d\tau,
\end{equation}
where the prime is used to denote the derivative with respect to time, $ '\equiv\partial_t $.
One can show \cite {Aminov} that the squared radius of adjoining sphere (in our case, a unit sphere)  is related to a curvature, torsion, and velocity modulus of an apex  by the formula
\begin{equation} \label {eq:13}
b^2=1 =\frac {1} {k^2} + (\frac {k \,'} {\mathrm {v} k^2 \kappa}) ^2.
\end{equation}
For the closed trajectories it is possible to introduce the integral characteristics:
 \begin{equation} \label {eq:14}
 C=\oint_{\Gamma} \vec {p} \, '\cdot d \vec {p},\,
 L=\oint_{\Gamma_1}\oint_{\Gamma_2}\frac{d \vec {p_1}'\cdot \,d \vec {p_2}'}{|\vec{ p_1}'-\vec{ p_2}'| },
 \end{equation}
  where $C$ is a circulation of vector $\vec {p} \, '$ along the closed path $\Gamma$.
 For two spins $L$ is a geometrical quantity (mutual induction ratio) in which the element of length $d\vec {p_1} ' \in \Gamma_1 $ is multiplied scalarwise on $d\vec {p_2} '\in \Gamma_2 $ and product is divided into the distance of these elements from each other. This parameter depends only on a configuration and a relative position of contours, as well as on a choice of a direction of positive bypass of each contour.
 \section {Rabi modulation}\label{sec:4}
   Taking into account the environment with parameters
  $ \gamma_2 = \gamma_1 =\gamma, R _{eqr} =0$,  in terms of the Bloch vector
  the Rabi model  \cite{IIRabi} is described by a matrix equation with an initial state
 $R (t=0) = (R_1 (t=0), R_2 (t=0), R_3 (t=0)) ^T $ (here $T$ indicates transposition)
 \begin{equation} \label {eq:15}
\partial_t R = M R,
\end{equation}
where the matrix looks as $ M = $ \begin {math} \bigl (
  \begin {smallmatrix}
  \gamma &-H & h \sin \omega t \\H & \gamma &-h \cos \omega t \\-h \sin \omega t & h \cos \omega t & \gamma
  \end {smallmatrix} \bigr)
\end {math}.
 \\
Let's introduce  vector $r=aR $ in which the orthogonal matrix is equal to

$a = $\begin {math} \bigl (
    \begin {smallmatrix}
  \cos \omega t & \sin \omega t & 0 \\
  -\sin \omega t & \cos \omega t & 0 \\
  0 & 0 & 1 \end {smallmatrix} \bigr)
\end {math}. The equation for $r $ takes the form
\begin{equation} \label {eq:16}
\partial_t r = \tilde {M} r
\end{equation}
with the constant matrix $ \tilde {M} = $ \begin {math} \bigl (
    \begin {smallmatrix}
  \gamma &-\delta & 0 \\
  \delta & \gamma &-h \\
  0 & h & \gamma
\end {smallmatrix} \bigr)
\end {math}.
Now the solution can be written as:
$R=a ^ {-1} e ^ {\tilde {M} t} R (t=0) $. For the  initial state
$R (t=0) = (\cos \varphi _0 \sin \theta _0,
\sin
\varphi _0 \sin \theta _0, \cos \theta _0) ^T $
%
for  $ \varphi _0 =\theta _0=0$ the solution becomes
%
\begin{eqnarray}\label {eq:17}
R_1&=&\frac {h e ^ {-\gamma t} } {\Omega ^2} (\Omega \sin \Omega t \sin \omega t + \delta
(1-\cos \Omega t) \cos \omega t),\\
R_2&=&\frac {h e ^ {-\gamma t} } {\Omega ^2} (\delta
(1-\cos \Omega t) \sin \omega t-\Omega \sin \Omega t \cos \omega t),\\
R_3&=&\frac {e ^ {-\gamma t}} {\Omega ^2} (\delta^2+h^2 \cos \Omega t),
\end{eqnarray}
%
where $ \delta=H-\omega $ and $ \Omega =\sqrt {\delta^2+h^2} $
is the  nonresonance Rabi frequency. \\
 The spin-flip probability  is equal to
   \begin {equation} \label {eq:20}
  P =\frac {1-R_3} {2}.
\end{equation}
This probability at a resonance $ \delta=H-\omega=0$ has
an oscillating behavior and also reaches a unity. At big
 detuning $ \delta \gg 1$ the spin-flip probability  tends to zero.
If the longitudinal field $H=0$, then the peak probability is equal to
1/2.\\
 \indent In the case of the elliptic field \eqref{eq:1}, 
  as the expansion
 of the Rabi model,
the solution in the elliptic field  is the following:
\begin{equation} \label {eq:21}
R_1 = e ^ {-\gamma t} \mathrm {sn} (\omega t|k) \sin h t, R_2 =
-e ^ {-\gamma t} \mathrm {cn} (\omega t|k) \sin h t, R_3=e ^ {-\gamma
t} \cos h t.
\end{equation}
In the model under consideration, the  Bloch vector components   decay in due course,
but the unit polarization vector $ \vec {p} = (p_1, p_2, p_3) $ does not depend on $ \gamma $.

\subsection{Visualization of dynamics in a circularly polarized field}
Let's give results for the explicit solution of the Rabi  model  for the initial
state $ \varphi _0 =\theta _0=0$.
In the circularly  polarized field $\overrightarrow {h (t)} = (h \cos \omega t, \, h \sin \omega t
, \, H) $
 the nutation and precession   velocities have the form
\begin{equation} \label {eq:22}
\theta \, ' = \frac {h^2 \Omega \sin \Omega t
   } {\sqrt {\Omega ^4-\left (\delta ^2+h^2 \cos
   \Omega t \right) ^2}}, \, \varphi \,' =
\frac {\omega \delta ^2 +\Omega ^2
   \delta + \omega \Omega ^2-\omega
   \left (\delta ^2-\Omega ^2\right)
   \cos \Omega t} {\delta ^2 +\Omega
   ^2 +\left (\Omega ^2-\delta ^2\right)
   \cos \Omega t}.
   \end{equation}
   \\
  \indent But in the case of  resonance,    the  nutation velocity   is the constant and the  precession velocity  is the constant too
\begin{equation} \label {eq:23}
\theta \,' (\delta=0) = h \text {sgn} (\sin h t), \,
\varphi \,' (\delta=0) = \omega
   \end{equation}
with the period $T=2 \pi/h $. \\
\indent The curvature, torsion, velocity, path length and circulation
 at the exact resonance  are expressed by the formulae
\begin{eqnarray}\label {eq:24}
k _ {res}&=&\frac {\sqrt {\left (h^2+3 \omega
   ^2\right) \left (8 h^4+4 \omega ^2
   h^2 +\omega ^4\right)-g (t)}} {\left (2 h^2 +\omega ^2-\omega
   ^2 \cos 2 h t\right) ^ {3/2}},\\
\kappa _ {res}&=&-\frac {4 h \omega \left (4 h^4+7
   \omega ^2 h^2 +\omega ^4 +\omega ^2
   \left (h^2-\omega ^2\right) \cos 2
   h t\right) \sin h
   t} {\left (h^2+3 \omega ^2\right)
   \left (8 h^4+4 \omega ^2 h^2 +\omega
   ^4\right)-g (t)
   },\\
\mathrm {v} _ {res}&=&\frac {1} {\sqrt {2}} \sqrt {2 h^2 +\omega ^2-\omega ^2 \cos 2 h   t},\\
s _ {res}&=&F (h t |-\frac {\omega^2} {h^2}),\,
C_ {res}=2 \pi  h+\frac{\pi  \omega ^2}{h},
\end{eqnarray}
 where $g(t) =4
    (\omega ^4-h^4+3 \omega ^2 h^2) \omega
   ^2\cos 2 h t + \left(\omega ^2-h^2\right)
   \omega ^4\cos 4 h t $, \\
      $F (\phi|m) = \int_0 ^\phi (1-m\sin^2\vartheta) ^ {1/2} d\vartheta $
   is the elliptic integral of the second kind. \\
\indent In the consistent Rabi field (cRf) Eq. \eqref {eq:1}, at the resonance, the angular velocities depend on the time and are equal to
\begin{equation} \label {eq:28}
\theta _ {cRf} \,' (\delta=0) = h \text {sgn} (\sin h t),
\varphi _ {cRf}\, ' (\delta=0) = \omega \text {dn} (\omega t|k).
\end{equation}
\section {Spin $S> 1/2$}\label{sec:5}
The qudit is characterized by the generalized Bloch vector,  the  dimension of  which is equal  to $ (2S+1) ^2-1$. The so developed scheme for a spin 1/2 is completely applicable for qudits,  for  the polarization vector  visualization and  the anisotropy influence on
polarization. For the qutrit, the 2D parametric time dependence of the  Bloch vector components,  was studied \cite{HoandChu,THoandSChu}. \\
\indent The Hamiltonian of the spin 1 (qutrit) in a  external variable magnetic field $ \overrightarrow {h (t)} $ taking into account anisotropy has the form \cite{finitequtritchain}
\begin{equation} \label {eq:29}
\hat {H}(h,Q,d) =h_i (t) S_i+Q (S_3^2-\frac {2} {3} E _ {3\times3}) +d (S_1^2-S_2^2),
\end{equation}
 where $S_i $ are the spin matrixes and  the anisotropy constants $Q, \, d $ take account of the contributions of the one-quantum and double-quantum transitions in qutrit \cite{finitequtritchain}.
The explicit form of the  spin matrices  and the set of equations, determining the qutrit evolution
are given in the paper \cite{finitequtritchain}. The exact solution in the circularly polarized field at the resonance without the influence of constant $d $  is given in the Appendix Eqs. \eqref{eq:31}-\eqref{eq:38}.
 \\
\indent
For the closed paths at $ \omega=0$ in the solution \eqref{eq:31}-\eqref {eq:38}
the requirement of the frequency  commensurability at $d=0$ looks as
$ m \sqrt{4 h^2+Q^2}  =n Q $,
 from which it follows that
 \begin{equation} \label{eq:30}
 h =\pm \sqrt {n^2-m^2} \frac {Q} {2 m},
 \end{equation}
 where $n, \, m $   are the integer numbers.
   It is possible to  show that the condition for the closed trajectory at $d\neq 0$ has the form
  $h +\sqrt {2} d =\pm \sqrt {n^2-m^2} \frac {Q} {2 m} $.
   \section {Numerical results}\label{sec:6}
\begin{figure}[h]
\includegraphics [width = 1.6 in] {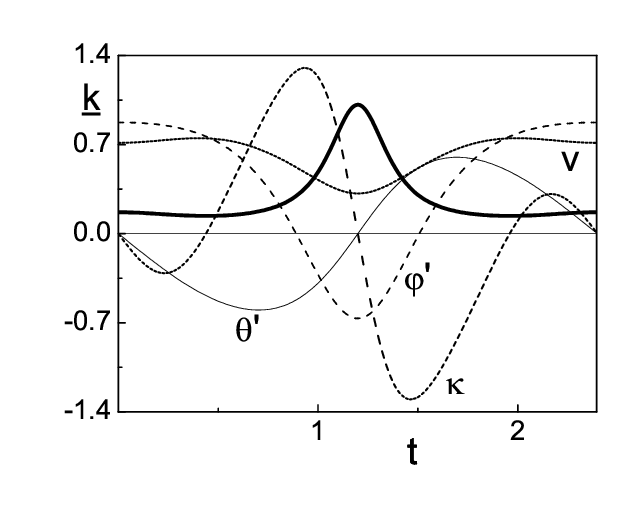}
\caption {\label {Fig_1_AllVisualization} Time dependencies of the curvature
$k $, torsion $ \kappa $,  velocity  $ \mathrm {v}, $
 precession and nutation velocities  $ \varphi ' $ , $ \theta ' $, for the
initial state  $ \theta_0 =\arccos\frac {1} {\sqrt {3}} $, \,
$ \varphi_0=0$ with the parameters of the circularly polarized field $ \omega=3$, \, $ H=0.45$,
\, $ h =-0.6$; \, $ \underline {k} =k/20$ during  $2 \pi/\Omega $.}
\end{figure}

\begin{figure}
 \includegraphics [width=1.6 in] {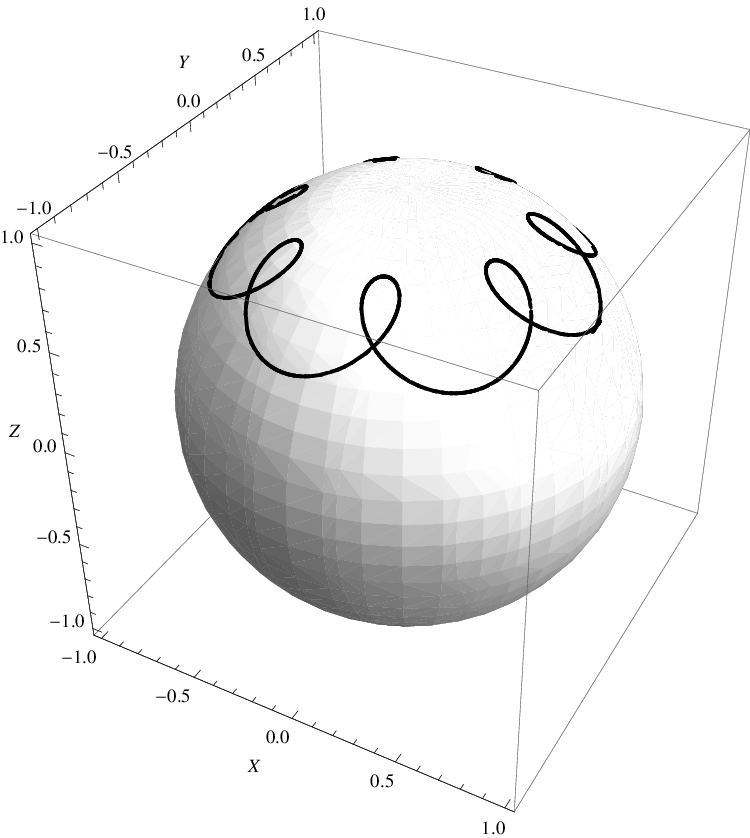}
 \caption {\label {Fig_2_loops}
 Mapping of the  qubit dynamics  on the  apex hodograph   (quantum interference of  precession and nutation) with the parameters as in Fig.~\ref {Fig_1_AllVisualization} for total time $T _ {all} =7 (2\pi/\Omega) $.
   The   spin-flip probability  oscillates in the limits of $ 0.06\leq P \leq 0.211 $. The path length equals $s=10.44. $}
\end{figure}

\begin{figure}
 \includegraphics [width=1.6 in] {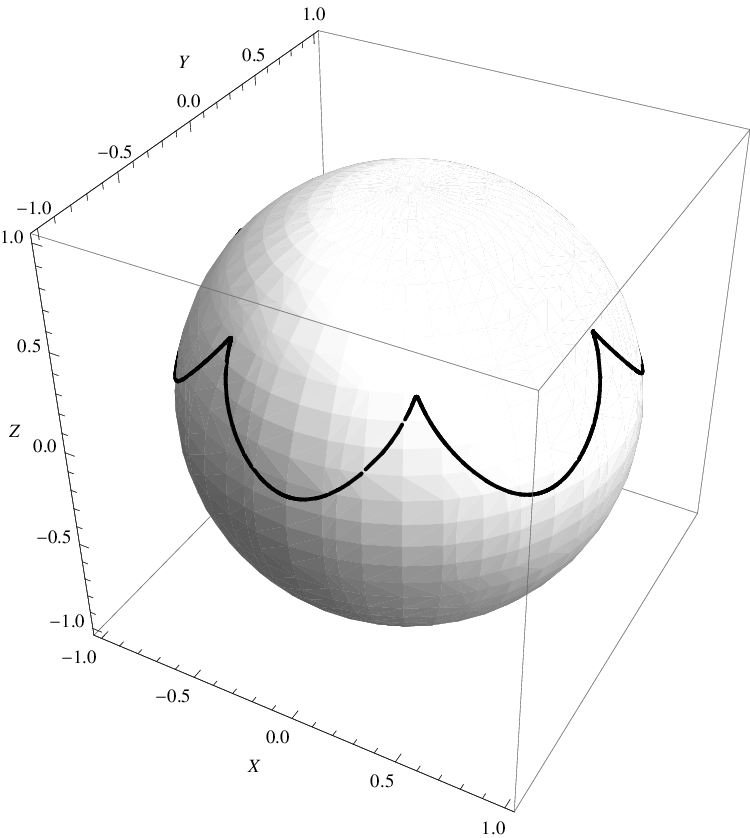}
 \caption {\label {Fig_3_cusps}
  Cusps  at
  $ \theta_0 =\arccos\frac {1} {\sqrt {3}}, \, \varphi_0 =\pi/4$;
       $ \omega=3, \, H=0.5, \, h=0.6$ during $6 (2\pi/\Omega) $;
     $0.19 \leq P \leq 0.4$;
    $ 0.015 \leq \mathrm {v} \leq 0.78$, $-0.02 \leq \varphi_t \leq 0.67$,
   $-0.6\leq \theta_t \leq  0.6$,
 $1 \leq k \leq 8500$, $-1550 \leq\kappa\leq1 550$, $ s=8.6$.}
\end{figure}

\begin{figure}
 \includegraphics [width=1.6 in] {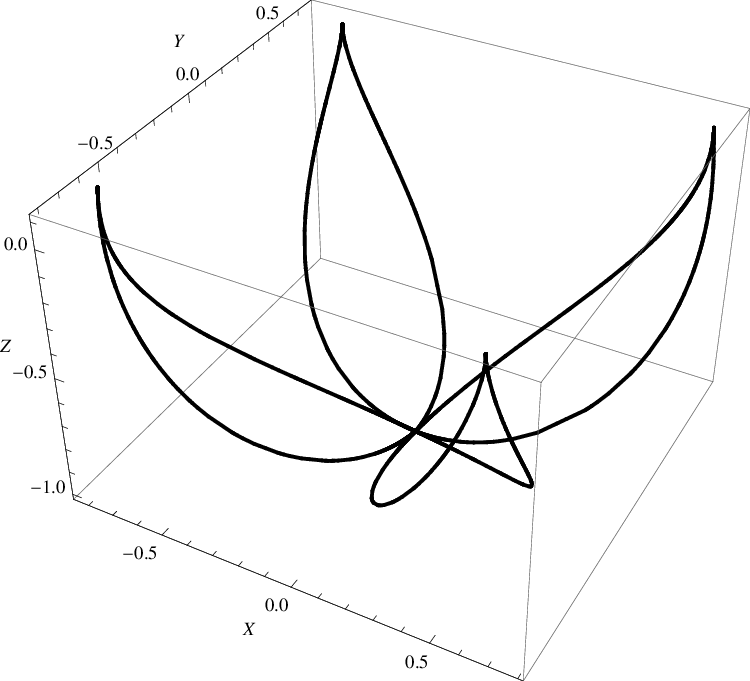}
 \caption {\label {Fig_4_rosa4}
    3D  parametric plot  of  the  qubit polarization vector  $\vec p (t) $ vs.  time.
  $ \theta_0 =\pi,
  \, \varphi_0=3 \pi/4$;
      $ \omega=0.5, \, H=0.05, \, h=0.5$ during $4 (2\pi/\Omega) $;
    $0.005 \leq \mathrm {v} \leq 0.5$,
 $1.5 \leq k \leq 25000$, $-20\leq\kappa\leq 20$,
     $-0.003\leq \varphi_t \leq 0.28$,
   $-0.5\leq \theta_t \leq 0.5$, $0.45 \leq P \leq 1$, $s=14$.}
\end{figure}

\begin{figure}
 \includegraphics [width=1.6 in] {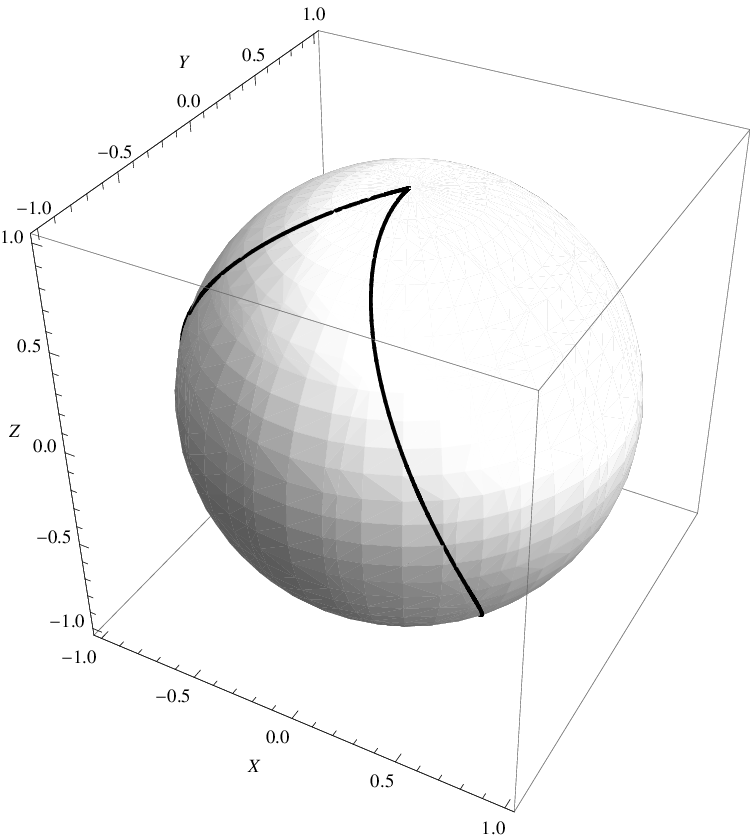}
 \caption {\label {Fig_5_resonance}
  Apex hodograph at a resonance
   for initial state $ \theta_0=0,
  \, \varphi_0 =\pi/4$;
    $ \omega= H=0.2, \, h=0.5$ for a period
 $T = 2\pi/h $;
    $0 \leq P \leq 1$,
  $0.5\leq \mathrm {v} \leq 0.54$, $1\leq k \leq 1.28$, \
    $-0.64\leq \kappa \leq 0.64$, $ s=6.53$,\,$ C_{res}=3.4$.}
\end{figure}

\begin{figure}
 \includegraphics [width=1.6 in] {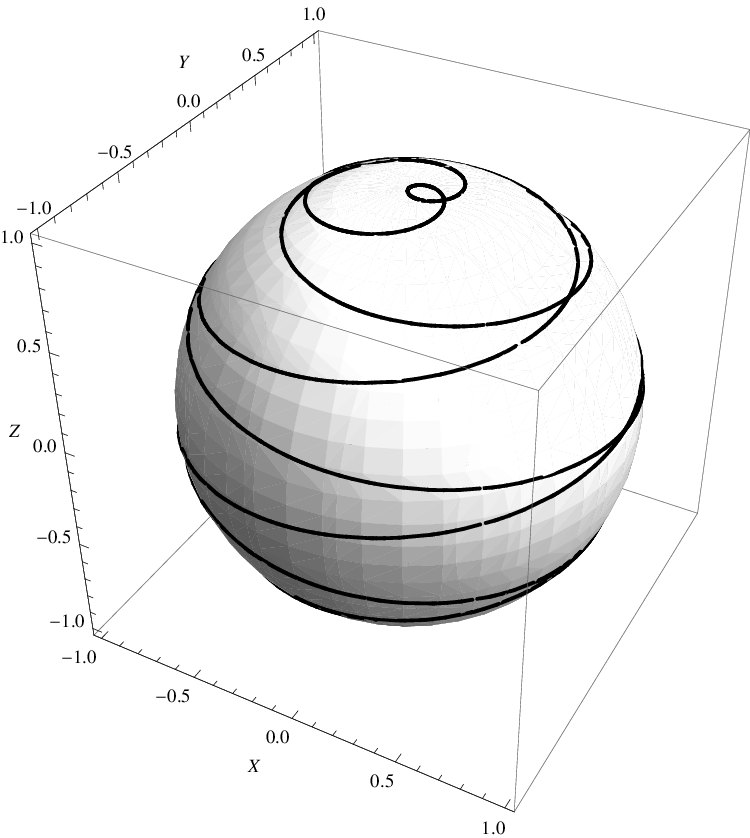}
 \caption {\label {Fig_6_resonance5505}
  Hodograph at the resonance
   for initial state  $ \theta_0=0,
  \, \varphi_0 =\pi/4$;
        $ \omega= H=5, \, h=0.5$, \, $T = 2\pi/h $;
     0$ \leq P \leq 1$,
      $0.47\leq \mathrm {v} \leq 5.02$, \, $1\leq k \leq 20$,
    $-0.55\leq \kappa \leq 0.55 $, \, $s=40,84$,\,$ C_{res}=160.14$.}
\end{figure}

\begin{figure}
 \includegraphics [width=1.6 in] {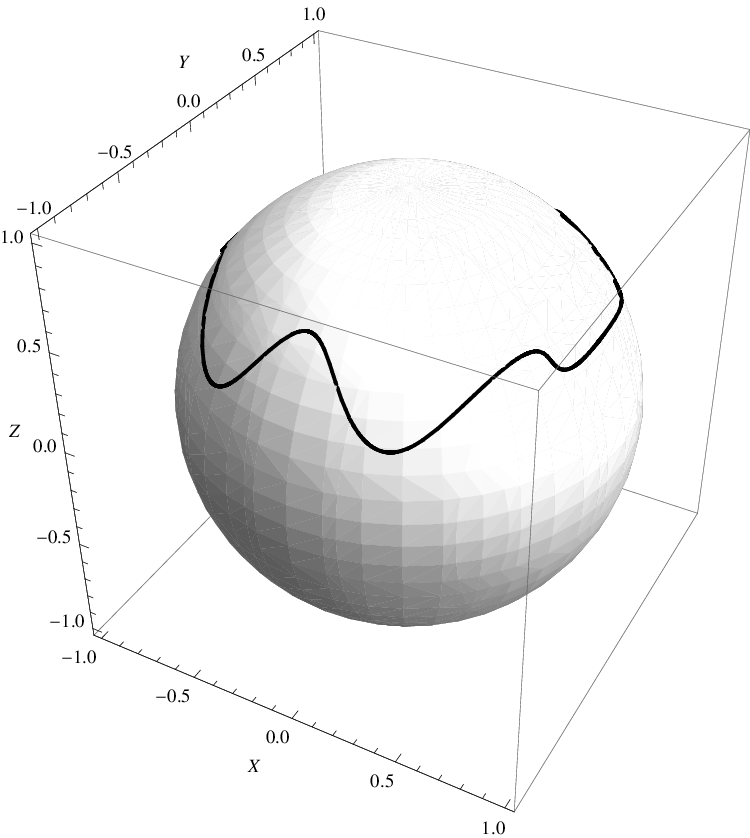}
 \caption {\label {Fig_7_h_2_0}
  Apex  qubit dynamics  in the
     linearly polarized field with the parameters as in Fig.~\ref {Fig_3_cusps}
     during  $14.66$;
    $0.14 \leq \mathrm {v} \leq 0.77$,
 $1 \leq k \leq 22$, $-200 \leq\kappa\leq 350$,
     $0.08 \leq \varphi_t \leq 0.87$,
   $-0.58\leq \theta_t \leq 0.6$, $0.14 \leq P \leq 0.3$, $s=6.44.$}
\end{figure}

\begin{figure}
 \includegraphics [width=1.6 in] {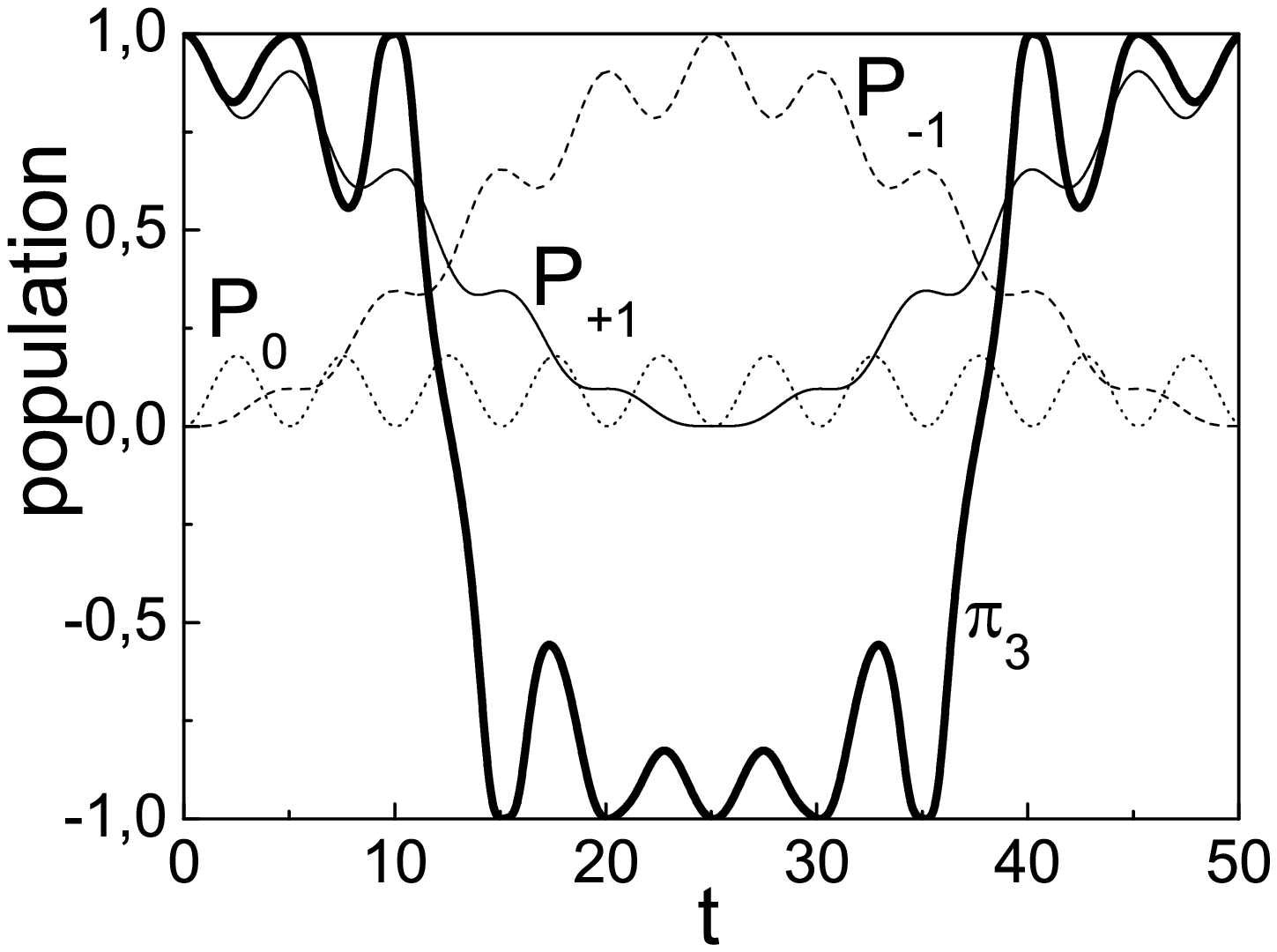}
 \caption {\label {Fig_8_qutrit_polarizationQ1}
  Populations Eqs. \eqref{eq:39} and  the third component
  of the  qutrit polarization vector $ \pi_3$
    with the model parameters (for $m=4, \, n=5$ in the  Eq.\eqref {eq:23})
    $ \omega=H=0, \, Q=1$, $h=3Q/8$ for the full time period $10 (2\pi/f). $
     }
\end{figure}

\begin{figure}
 \includegraphics [width=1.6 in] {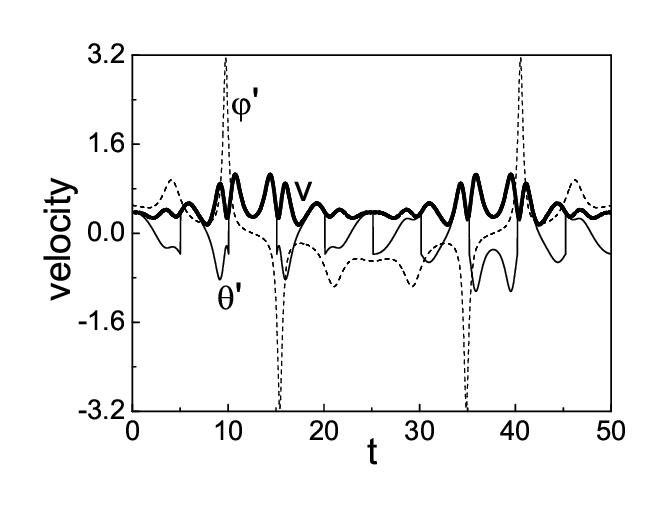}
 \caption {\label {Fig_9_velocity}
    Time dependence of the  velocities  of the qutrit apex  $ \mathrm {v}, $
 $ \varphi ' $ and $ \theta ' $
     with the model parameters as in a Fig.~\ref {Fig_8_qutrit_polarizationQ1}.
     }
\end{figure}

\begin{figure}
 \includegraphics [width=1.6 in] {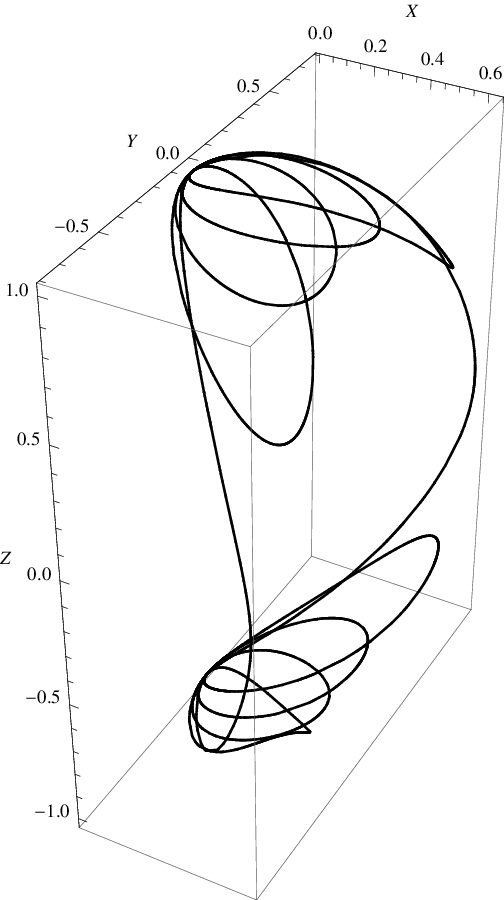}
 \caption {\label {Fig_10_QUTRIT_Q_1}
   3D parametric plot  of the qitrit polarization vector $ \vec\pi(t)$
  with the parameters as in  Fig.~\ref {Fig_8_qutrit_polarizationQ1}. The torsion is small
  $-10 ^ {-26} <\kappa <10 ^ {-26} $  and also changes its sign  28 times over the full period, while  the curvature varies within 0.005$ \leq k\leq 29$;
   $s=22.13. $}
\end{figure}

\begin{figure}
 \includegraphics [width=1.6 in]{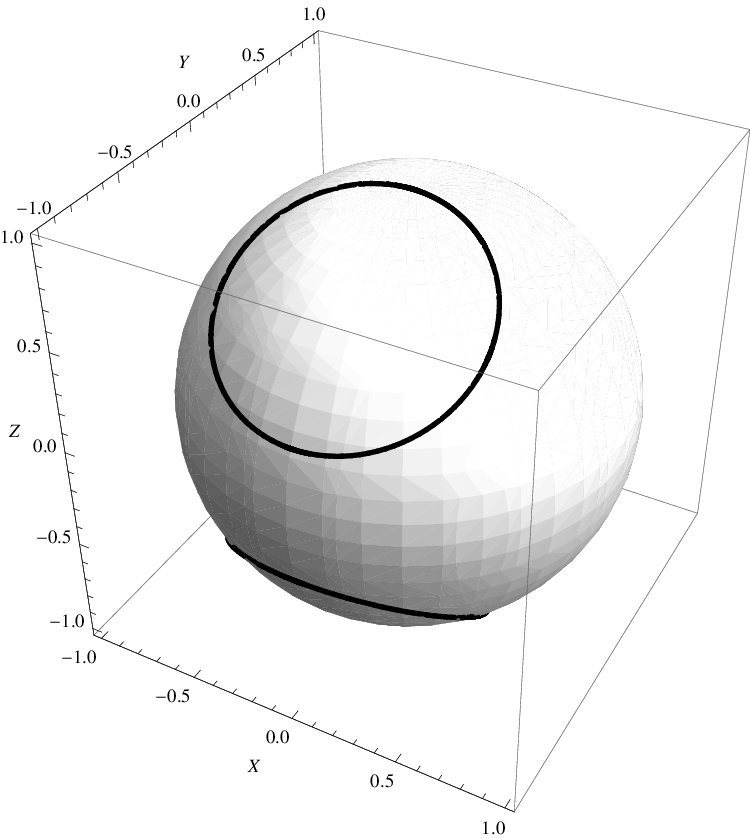}
 \caption {\label {Fig_11_2Qutrit}
   Hodographs for 2 qitrit polarization vectors on the sphere  over the time  25.07; $J=0.01,  h_1=\bar{h}_1=0.15,  h_3=-\bar{h}_3=0.2$;\, circulations:   $C_{up}=0.53,\,C_{down}=-0.81.$}
\end{figure}

 \textit{Qubit}. In the circularly  polarized field in
Fig.~\ref{Fig_1_AllVisualization} the
 curvature, torsion and velocities are presented, which for descriptive reasons
are combined by introducing a scale factor (for the curvature).
 It becomes clear  that when the  torsion $ \kappa $ changes its sign from plus to minus,  the velocity modulus  of the apex  $ \mathrm {v} \equiv |\vec {p}' | $ decreases,  the  precession velocity is minimum, the   nutation velocity   changes its sign from  minus to plus and  the curvature $k $ increases.
 The interaction of the effects of precession and  nutation
 is geometrically described in  Fig.~\ref{Fig_2_loops}.
The occurrence of loops (self-superposition) follows from the fact that the precession velocity
 on the upper parallel  is opposite to the velocity on the lower parallel. \\
 \indent At the cusps, as it is obvious in Fig.~\ref{Fig_3_cusps},
there is a sharp transition from some geometrical characteristics of the hodograph to others (that is, linking trajectories
 with different $k, \kappa $). In the transition vicinity
 the velocity $\mathrm {v}$ sharply decreases \cite{BenentiSiccardiStriniPhysRevA},
  the curvature and torsion sharply
 increase and the  precession and  nutation velocities tend to zero. \\
   \indent If motion begins from $ \theta=0$ or
   $ \theta =\pi $, then the upper or accordingly the lower parallel of the sphere
  degenerates in the point.  Fig.~\ref{Fig_4_rosa4} presents dynamics,
   where in the vicinity of the cusps the velocity  $ \mathrm {v} $ sharply
  decreases, the curvature and torsion sharply increase,
  and the precession and nutation velocities actually transit through
  zero.    In all the cases in the cusps the qubit energy
    $E _ {Rabi} = \mathrm {Tr \,} (\rho \hat {H}) $ has its local maximum.
    \\
\indent At the coincidence of the external field frequency  and the eigenfrequency  there is the resonance in the system.
  On the sphere in Fig.~\ref{Fig_5_resonance} it is visible that at very small
   frequencies which correspond to the small   precession velocity
 Eq. \eqref{eq:23}, on the closed apex trajectories (with beginning on the	
 northern pole towards the southern one and  termination on the northern  pole
  for one period), which is a purely periodic driving with a period
   $T=2\pi/h $), there are no self-intersections.
  At the constant Rabi  frequency $h $,
 (see  Fig.~\ref{Fig_6_resonance5505}), the increase of the frequency $ \omega $  causes the increase of the path length.
   Let's emphasize that the trajectory is not flat, but tri-dimensional. This is visible along the path, which is more than 2$ \pi $, and by means of another criteria it is possible to show that for the plane curve torsion is equal to zero in each point. The  whole class of closed trajectories satisfies the requirement for the closed curve which  is determinated by the formula
$h =\pm\sqrt {n^2 \omega
   ^2/m^2 - (H-\omega) ^2} $. \\
 \indent As it is known \cite{BlochSiegert}, in the linearly polarized field
 $ \overrightarrow {h (t)} = (h \cos \omega t, \, 0, \, H) $, the resonance happens at  $ \omega \neq H $, (the Bloch-Siegert resonance frequency shift).
 The visualization for such modulation was partially considered in the
 paper \cite{BenentiSiccardiStriniPhysRevA}. We would like only to add that in the linearly polarized field instead of loops or cusps, the hodograph obtained from the numerical solution, at parameters such as in Fig.~\ref{Fig_3_cusps} is more smooth (see  Fig.~\ref {Fig_7_h_2_0}) as the   precession velocity  does not change its sign and the curvature and torsion in such field are much less than in the rotating field. \\
 \indent \textit{Qutrit}.
We present some numerical results describing qutrit dynamics in the field $ \overrightarrow {h (t)} = (h, \, 0, \, 0) $ for $d=0$. The smooth closed trajectory
goes through the both  poles that means that there is a two-photon transition.
In more detail, a flip of the third component of polarization vector $ \pi_3$ occurs and the population of the lowest level $P _ {-1} $ is equal to unity over the time ($5 (2\pi/f) $).
In the vicinity of the southern pole, the precession velocity takes its minimum value and the nutation velocity changes its sign from plus to minus.
Fig. ~\ref {Fig_8_qutrit_polarizationQ1} shows both dynamics of the level populations with spin projections (1,0,-1) and third projection of the polarization vector $ \pi_3$. Fig.~\ref {Fig_9_velocity} presents the dynamics of velocities.
A comparison with the hodograph, given in Fig.~\ref{Fig_10_QUTRIT_Q_1}, shows that the precession corresponds to small oscillations of $ \pi_3$  (two orbits on the sphere over the time $2 (2\pi/f) $). Then the sharp nutational change of $ \pi_3$ (the smooth curve) occurs during  $ (2\pi/f) $. Next four precessional orbits go through the south pole of the sphere during  $4 (2\pi/f) $). Further sharp nutational change of $ \pi_3$ (the smooth curve) takes place during $ (2\pi/f) $. The full period $10 (2\pi/f) $ is completed by two precessional orbits on the northern pole of the sphere over the time $2 (2\pi/f) $).
In other words, there is a quasitrapping $ \pi_3$ on the poles. All the velocities $\mathrm {v}$,  $\varphi' $, $ \theta' $ grow with increasing anisotropy and the periods of "hodograph components" decrease.
However, the shape of hodograph does not depend on the anisotropy parameter at positive $Q$. The torsion $ \kappa $ is small and repeatedly changes its sign during the full period. In fact, such a hodograph is the Lissajous figure on the tri-dimensional sphere. \\
 \indent The influence of the anisotropy constant $d$ caused by the double-quantum transitions leads to the increase of the transition frequency.
When $d=0.1$, the frequency of the double-quantum transitions is increased approximately twice, due to reduction of the number of the precessional orbits on the poles.\\
\indent \textit{2 qutrits}.  Having solved equation system for two qutrits with the Hamiltonian $\hat {H}_2=\hat {H}(h,Q,d)\bigotimes E_{3\times3}+E_{3\times3}\bigotimes\hat {H}(\bar{h},\bar{Q},\bar{d})+J \vec{S}\bigotimes \vec{S}$ numerically  \cite{finitequtritchain} in the constant magnetic field with the isotropic interaction and the initial state of the antiparallel polarization vectors pointed at poles of the sphere (see Fig. ~\ref {Fig_11_2Qutrit}), we find the mutual induction ratio dependence on an exchange constant: $L (J =-0.01) =-5.61, \, L (J=0) =-5.81, \, L (J=0.01) =-5.93$.\\
\indent The results obtained well exhibit quantum interference of the precession and nutation components of the motion of the polarization vector in resonance and nonresonance conditions.
 \section {Conclusion}\label{7}
A system of closed equations describing dynamics of the Bloch vector components for a qubit in an arbitrary time-dependent external magnetic field is derived.
On the basis of analytical solutions we propose a scheme of the qubit dynamics visualization. This is realized by mapping the solution on a tri-dimensional parametric hodograph of an apex of polarization vector on a sphere. The numerical results for the qubit dynamics in a circularly polarized field are presented for different initial states.  Such representation and its numerical validation are important also in view of a deeper insight in the relation between classical and quantum randomness, opening unexpected, possible applications of the mathematical formalism of quantum mechanics in other domains of science. The proposed scheme of visualization is also applicable for 3D mapping the analytical or numerical solution of  coupled qudits.\\
\indent We believe that this study will be useful since 3D visualization of polarization vectors can be applicable to a double magnetic resonance and manipulations of quantum bits. The visualization may have potential applications in magnetometery and quantum information processing.\\
\clearpage
\appendix
\section*{Appendix} \label {A}
  The exact solution for the qutrit,  taking into account only the anisotropy $Q $ at the start from the northern pole
 $ \rho _ {qutrit} (t=0) = $ \begin {math} \bigl (
    \begin {smallmatrix}
  1 & 0 & 0 \\
  0 & 0 & 0 \\
  0 & 0 & 0 \end {smallmatrix} \bigr)
\end {math} in the  circularly  polarized magnetic field
 \cite {IIRabi} at the resonance $ \omega=H $ has the form
 \begin{equation} \label {eq:31}
 q_1 =
\frac {\sqrt {6} h \sin \frac {f t} {2}} {f^2}
  (f \cos \frac {Q t} {2} \sin
    \omega t+Q \sin \frac {f t} {2} \cos
   \omega t),
\end{equation}
\begin{equation} \label {eq:32}
q_2 =\frac {\sqrt {6} h \sin \frac {f t} {2}} {f^2}
   (Q \sin \frac {f t} {2} \sin
   \omega t-f \cos \frac {Q t} {2} \cos
    \omega t),
\end{equation}
\begin{equation} \label {eq:33}
q_3 = \sqrt {\frac {3} {2}} (\frac {Q \sin \frac {f
   t} {2} \sin \frac {Q
   t} {2}} {f} + \cos \frac {f t} {2}
   \cos \frac {Q t} {2}),
   \end{equation}
   \begin{equation} \label {eq:34}
q_4 =
\frac {1} {f^2} \sqrt {\frac {3} {2}}
(-2 h^2 \sin
   ^2\frac {f t} {2}
   \sin 2 \omega t + \left (f^2 \cos
   \frac {f t} {2} \sin
   \frac {Q t} {2}-f Q
   \sin \frac {f t} {2}
   \cos \frac {Q
   t} {2} \right) \cos 2 \omega t)
  ,
   \end{equation}
   \begin{equation} \label {eq:35}
q_5 =
\frac {\sqrt {6} h \sin \frac {f t} {2}} {f}
   (\sin \frac {Q t} {2} \sin
   \omega t-\cos \frac {f t} {2} \cos
   \omega t),
   \end{equation}
   \begin{equation} \label {eq:36}
   q_6 =\frac {h^2+Q^2+3 h^2 \cos f t} {\sqrt {2} f^2},
   \end{equation}
   \begin{equation} \label {eq:37}
   q_7 =
\frac {\sqrt {6} h \sin \frac {f t} {2}} {f}
   (\cos \frac {f t} {2} \sin
   \omega t +\sin \frac {Q t} {2} \cos
   \omega t),
   \end{equation}
   \begin{equation} \label {eq:38}
   q_8 =
\frac {1} {f^2} \sqrt {\frac {3} {2}}
(f
    (Q \sin
   \frac {f t} {2} \cos
   \frac {Q t} {2}-f
   \cos \frac {f t} {2}
   \sin \frac {Q
   t} {2}) \sin 2 \omega t-2 h^2 \sin
   ^2\frac {f t} {2}
   \cos 2 \omega t),
   \end{equation}
   where $ q_1, \, q_2, \, q_3$ are the  spin components, \, $f =\sqrt {4 h^2+Q^2}. $  The population levels in the qitrit are equal to
\begin{equation} \label {eq:39}
P _ {+ 1} = \frac {1} {6} (2 +\sqrt {6} q_3 +\sqrt {2} q_6), \,
P _ {0} = \frac {1} {3} (1-\sqrt {2} q_6), \,
P _ {-1} = \frac {1} {6} (2-\sqrt {6} q_3 +\sqrt {2} q_6).
\end{equation}
The   qutrit polarization vector   is equal to
$ \vec\pi(t)=(q_1 ,\,q_2,\,q_3)/N $, where
$N =\sqrt {q_1^2+q_2^2+q_3^2} $.
\indent The author  wishes to thank  I. V. Tanatarov and A. S. Peletminskii  for useful comments.
%









\end{document}